# The periodic table of elementary particles

Ding-Yu Chung

P.O. Box 180661, Utica, Michigan 48318, USA

All leptons, quarks, and gauge bosons can be placed in the periodic table of elementary particles. The periodic table is derived from dualities of string theory and a Kaluza-Klein substructure for the six extra spatial dimensions. As a molecule is the composite of atoms with chemical bonds, a hadron is the composite of elementary particles with hadronic bonds. The masses of elementary particles and hadrons can be calculated using the periodic table with only four known constants: the number of the extra spatial dimensions in the superstring, the mass of electron, the mass of Z°, and $\alpha_e$. The calculated masses are in good agreement with the observed values. For examples, the calculated masses for the top quark, neutron, and pion are 176.5 GeV, 939.54MeV, and 135.01MeV in excellent agreement with the observed masses, 176 ± 13 GeV, 939.57 MeV, and 134.98 MeV, respectively. The masses of 110 hadrons are calculated. The overall average difference between the calculated masses and the observed masses for all hadrons is 0.29 MeV. The periodic table of elementary particles provides the most comprehensive explanation and calculation for the masses of elementary particles and hadrons.

## 1. Introduction

The mass hierarchy of elementary particles is one of the most difficult problems in particle physics. There are various approaches to solve the problem. In this paper, the approach is to use a periodic table. The properties of atoms can be expressed by the periodic relationships in the periodic table of the elements. The periodic table of the elements is based on the atomic orbital structure. Do elementary particles possess orbital structure? In this paper, the orbital structure for elementary particles is constructed from the superstring. The extra dimensions in the superstring become the orbits with the Kaluza-Klein substructure. In the Kaluza's model of five-dimensional spacetime, the fifth dimension is considered to be a one-dimensional circle associated with every point in ordinary flat four-dimensional spacetime. Similarly, in the Kaluza-Klein substructure for the extra spatial dimensions, the fifth dimension is a one-dimensional circle associated with every point in ordinary flat four-dimensional spacetime, the sixth dimension circles the circle of the fifth dimension, and in the same way, every higher dimension circles the circle of the lower dimension. The energy is assumed to be distributed differently in different dimensions in such way that the energy increases with increased number of spacetime dimensions. This Kaluza-Klein substructure is derived from the duality relations for the superstring. These duality relations provide the explanation for the co-existence of leptons and quarks, the structure of the dimensional orbits, and the compositions of elementary particles.

All elementary particles are placed in the periodic table of elementary particles [1]. As a molecule is the composite of atoms with chemical bonds, a hadron is the composite of elementary particles with hadronic bonds. The masses of elementary particles and hadrons can be calculated using only four known constants: the number of the extra spatial dimensions in the superstring, the mass of electron, the mass of Z°, and $\alpha_e$. The calculated masses of elementary particles and hadrons are in good agreement with the observed values. For examples, the calculated masses for the top quark, neutron, and pion are 176.5 GeV, 939.54MeV, and 135.01MeV in excellent agreement with the observed masses, 176 ± 13 GeV [2], 939.57 MeV, and 134.98 MeV, respectively.

## 2. Dualities

A membrane can be described as two dimensional object that moves in an eleven dimensional space-time [3]. This membrane can be converted into the ten dimensional superstring with the extra dimension curled into a circle to become a closed superstring. If the membrane is split into two membranes which are then converted into two superstrings connected by the extra dimensions. The membrane becomes a dual superstring. It is proposed that the dual superstring is the lepton-quark dual superstring consisting lepton superstring and quark superstring as shown in Fig. 1.

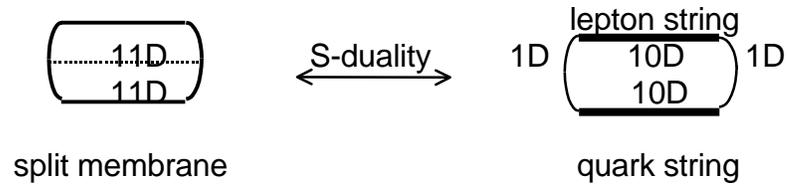

Fig. 1. S-duality

There is S-duality between the split membrane and the dual superstring. The membrane interacts strongly at the high energy, when the dual lepton-quark superstring interacts weakly. The four dimensional version of S-duality is Montonen-Olive duality [4] where the magnetic dipole is split into magnetic monopoles, and electrons and quarks arise as solitons.

M. Duff and J. Lu [5] suggested the U-duality between the four dimensional space-time solitonic string in six dimensional space-time and the fundamental ten dimensional superstring. In the same way, the dual lepton-quark superstring can also have two four dimensional space-time solitonic strings in two seven dimensional space-time. There is U-duality between the dual solitonic string and the dual lepton-quark superstring as shown in Fig. 2.

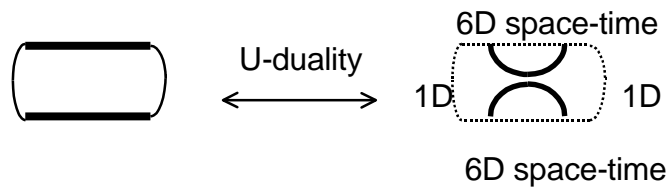

Fig. 2. U-duality

The Type II dual string can be compacified into dual D-branes in the dual 7D tori for the dual seven extra dimensions to form orbitfolds [6]. It is proposed that the energy associated with each extra dimension increases incrementally from the fifth to the eleventh dimension as shown in Fig. 3.

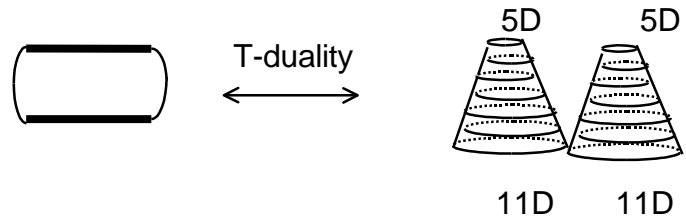

Fig. 3. T-duality

This process generates the Kaluza-Klein substructure for the dual tori. In the Kaluza-Klein substructure for the seven extra spatial dimensions, the fifth dimensional orbit is a one-dimensional circle associated with every point in ordinary flat four-dimensional space-time, the sixth dimensional orbit circles the fifth dimensional orbit, and in the same way, every higher extra dimensional orbit with higher energy circles the lower dimensional orbit with lower energy.

Each extra space-time dimension can be described by a fermion and a boson. The masses of fermion and its boson partner are not the same. This supersymmetry breaking is in the form of a energy hierarchy with increasing energies from the dimension five to the dimension eleven as

$F_5$ $B_5$ $F_6$ $B_6$ $F_7$ $B_7$ $F_8$ $B_8$ $F_9$ $B_9$ $F_{10}$ $B_{10}$ $F_{11}$ $B_{11}$

where B and F are boson and fermion in each spacetime dimension. The probability to transforming a fermion into its boson partner in the adjacent dimension is same as the fine structure constant, $\alpha$, the probability of a fermion emitting or absorbing a boson. The probability to transforming a boson into its fermion partner in the same dimension is also the fine structure constant, $\alpha$. This hierarchy can be expressed in term of the dimension number, D,

$$E_{D-1,B} = E_{D,F}\, \alpha_{D,F} \;, \tag{1}$$

$$E_{D,F} = E_{D,B}\, \alpha_{D,B} \;, \tag{2}$$

where $E_{D,B}$ and $E_{D,F}$ are the energies for a boson and a fermion, respectively, and $\alpha_{D,B}$ or $\alpha_{D,F}$ is the fine structure constant, which is the ratio between the energies of a boson and its fermionic partner. All fermions and bosons are related by the order $1/\alpha$. (In some mechanism for the dynamical supersymmetry breaking, the effects of order $\exp(-4\pi^2/g^2)$ where g is some small coupling, give rise to large mass hierarchies [7].) Assuming $\alpha_{D,B} = \alpha_{D,F}$, the relation between the bosons in the adjacent dimensions, then, can be expressed in term of the dimension number, D,

$$E_{D-1,B} = E_{D,B}\, \alpha^2{}_D \;, \tag{3}$$

or

$$E_{D,B} = \frac{E_{D-1,B}}{\alpha^2{}_D} \;, \tag{4}$$

where D= 6 to 11, and $E_{5,B}$ and $E_{11,B}$ are the energies for the dimension five and the dimension eleven, respectively.

The lowest energy is the Coulombic field, $E_{5,B}$

$$E_{5,B} = \alpha\, M_{6,F}$$

$$= \alpha\, M_e \;, \tag{5}$$

where $M_e$ is the rest energy of electron, and $\alpha = \alpha_e$, the fine structure constant for the magnetic field. The bosons generated are called "dimensional bosons" or "$B_D$". Using only $\alpha_e$, the mass of electron, the mass of $Z^0$, and the number of extra dimensions (seven), the masses of $B_D$ as the gauge boson can be calculated as shown in Table 1.

**Table 1.** The Energies of The Dimensional Bosons
$B_D$ = dimensional boson, $\alpha = \alpha_e$

| $B_D$ | $E_D$ | GeV | Gauge Boson | Interaction |
|---|---|---|---|---|
| $B_5$ | $M_e \alpha$ | $3.7 \times 10^{-6}$ | A | electromagnetic |
| $B_6$ | $M_e/\alpha$ | $7 \times 10^{-2}$ | $\pi_{1/2}$ | strong |
| $B_7$ | $E_6/\alpha_w^2 \cos\theta_w$ | 91.177 | $Z_L^0$ | weak (left) |
| $B_8$ | $E_7/\alpha^2$ | $1.7 \times 10^6$ | $X_R$ | CP (right) nonconservation |
| $B_9$ | $E_8/\alpha^2$ | $3.2 \times 10^{10}$ | $X_L$ | CP (left) nonconservation |
| $B_{10}$ | $E_9/\alpha^2$ | $6.0 \times 10^{14}$ | $Z_R^0$ | weak (right) |
| $B_{11}$ | $E_{10}/\alpha^2$ | $1.1 \times 10^{19}$ | | |

In Table 1, $\alpha_w$ is not same as $\alpha$ of the rest, because there is symmetry group mixing between $B_5$ and $B_7$ as the symmetry mixing in the standard theory of the electroweak interaction, and $\sin\theta_w$ is not equal to 1. As shown latter, $B_5$, $B_6$, $B_7$, $B_8$, $B_9$, and $B_{10}$ are A (massless photon), $\pi_{1/2}$, $Z_L^0$, $X_R$, $X_L$, and $Z_R^0$, respectively, responsible for the electromagnetic field, the strong interaction, the weak (left handed) interaction, the CP (right handed) nonconservation, the CP (left handed) nonconservation, and the P (right handed) nonconservation, respectively. The calculated value for $\theta_w$ is $29.69^0$ in good agreement with $28.7^0$ for the observed value of $\theta_w$ [8]. The total energy of the eleven dimensional superstring is the sum of all energies of both fermions and bosons. The calculated total energy is $1.1 \times 10^{19}$ GeV in good agreement with the Planck mass, $1.2 \times 10^{19}$ GeV, as the total energy of the superstring. As shown later, the calculated masses of all gauge bosons are also in good agreement with the observed values. Most importantly, the calculation shows that exactly seven extra dimensions are needed for all fundamental interactions.

In the Kaluza-Klein substructure, the bosons and the fermions in the dimensions > 5 gain masses in the four dimensional spacetime by combining with scalar Higgs bosons in the four dimensional spacetime. Without combining with scalar Higgs bosons, neutrino and photon remain massless. Graviton, the massless boson from the vibration of the superstring, remains massless.

### 3. The Internal Symmetries And The Interactions

All six non-gravitational dimensional bosons are represented by the internal symmetry groups, consisting of two sets of symmetry groups, U(1), U(1), and SU(2) with the left-right symmetry. Each $B_D$ is represented by an internal symmetry group as follows.

$B_5$: U(1), $B_6$: U(1), $B_7$: $SU(2)_L$, $B_8$: $U(1)_R$, $B_9$: $U(1)_L$, $B_{10}$: $SU(2)_R$

The additional symmetry group, U(1) X SU(2)$_L$, is formed by the "mixing" of U(1) in B$_5$ and SU(2)$_L$ in B$_7$. This mixing is same as in the standard theory of the electroweak interaction.

As in the standard theory to the electroweak interaction, the boson mixing of U(1) and SU(2)$_L$ is to create electric charge and to generate the bosons for leptons and quarks by combining isospin and hypercharge. The hypercharges for both e$^+$ and ν are 1, while for both u and d quarks, they are 1/3 [9]. The electric charges for e$^+$ and ν are 1 and 0, respectively, while for u and d quarks, they are 2/3 and -1/3, respectively.

There is no strong interaction from the internal symmetry for the dimensional bosons, originally. The strong interaction is generated by "leptonization" [10], which means quarks have to behave like leptons. Leptons have integer electric charges and hypercharges, while quarks have fractional electric charges and hypercharges. The leptonization is to make quarks behave like leptons in terms of "apparent" integer electric charges and hypercharges. Obviously, the result of such a leptonization is to create the strong interaction, which binds quarks together in order to make quarks to have the same "apparent" integer electric charges and hypercharges as leptons. There are two parts for this strong interaction: the first part is for the charge, and the second part is for the hypercharge. The first part of the strong interaction allows the combination of a quark and an antiquark in a particle, so there is no fractional electric charge. It involves the conversion of $\pi_{1/2}$ boson in B$_6$ into the electrically chargeless meson field by combining two $\pi_{1/2}$, analogous to the combination of $e^+$ and $e^-$ fields, so the meson field becomes chargeless. (The mass of $\pi$, 135 MeV, is twice of the mass of a half-pion boson, 70 MeV, minus the binding energy. $\pi_{1/2}$ becomes pseudoscalar up or down quark in pion.) In the meson field, no fractional charge of quark can appear. The second part of the strong interaction is to combine three quarks in a particle, so there is no fractional hypercharge. It involves the conversion of $B_5$ ($\pi_{1/2}$) into the gluon field with three colors. The number of colors (three) in the gluon field is equal to the ratio between the lepton hypercharge and quark hypercharge. There are three $\pi_{1/2}$ in the gluon field, and at any time, only one of the three colors appears in a quark. Quarks appear only when there is a combination of all three colors or color-anticolor. In the gluon field, no fractional hypercharge of quarks can appear. By combining both of the meson field and the gluon field, the strong interaction is the three-color gluon field based on the chargeless vector meson field from the combination of two $\pi_{1/2}$'s. The total number of $\pi_{1/2}$ is 6, so the fine structure constant, $\alpha_s$, for the strong force is

$$\begin{aligned}\alpha_S &= 6\alpha \ e^1 \\ &= 0.119\end{aligned} \qquad (6)$$

which is in a good agreement with the observed value, 0.124 [11].

The dimensional boson, $B_8$, is a CP violating boson, because $B_8$ is assumed to have the CP-violating $U(1)_R$ symmetry. The ratio of the force constants between the CP-invariant $W_L$ in $B_8$ and the CP-violating $X_R$ in $B_8$ is

$$\frac{G_8}{G_7} = \frac{\alpha \, E_7^2 \, \cos^2 \Theta_W}{\alpha_W \, E_8^2} \quad (7)$$

$$= 5.3 \times 10^{-10} \;\;,$$

which is in the same order as the ratio of the force constants between the CP-invariant weak interaction and the CP-violating interaction with $|\Delta S| = 2$.

The dimensional boson, $B_9$ ($X_L$), has the CP-violating $U(1)_L$ symmetry. The lepton ($l_9$) and the quark ($q_9$) are outside of the three families for leptons and quarks, so baryons in dimension nine do not have to have the baryon number conservation. The baryon which does not conserve baryon number has the baryon number of zero. The combination of the CP-nonconservation and the baryon number of zero leads to the baryon, $\bar{p}_9^{\,-}$, with the baryon number of zero and the baryon, $p_9^{\,+}$, with the baryon number of 1. The decay of $\bar{p}_9^{\,-}$ is as follows.

$$\bar{p}_9^{\,-} \rightarrow \bar{l}_9^{\,-} \; \bar{l}_9^{\,\circ}$$

The combination this $\bar{p}_9^{\,-}$ and $p_9^+$ as well as leptons, $l_9^{\,\circ} \; \bar{l}_9^{\,\circ}$ results in

$$p_9^{\,+} \; \bar{p}_9^{\,-} \; + l_9^{\,\circ} \; \bar{l}_9^{\,\circ} \rightarrow \; p_9^{\,+} + l_9^{\,-} \; \bar{l}_9^{\,\circ} + l_9^{\,\circ} \; \bar{l}_9^{\,\circ}$$

$$\rightarrow \; p_9^{\,+} + l_9^{\,-} + \text{radiation} + \bar{l}_9^{\,\circ}$$

$$\rightarrow \; n_9^{\,\circ} + \text{radiation}$$

Consequently, excess baryons, $n_9^{\,\circ}$, are generated in the dimension nine. These excess baryons, $n_9^{\,\circ}$, become the predecessors of excess neutrons in the low energy level.

The ratio of force constants between $X_R$ with CP-independent baryon number and $X_L$ with CP-dependent baryon number is

$$\frac{G_9}{G_8} = \frac{\alpha E_8^2}{\alpha E_9^2} \qquad (8)$$
$$= 2.8 \times 10^{-9} \quad,$$

which is in the same order as the observed ratio of the numbers between the left handed baryons (proton or neutron) and photons in the universe.

The direct transition of the superstring to $q_9$ without going through the last two extra dimensions ($B_{11}$, $F_{11}$, $B_{10}$, and $F_{10}$) may account for the inflation in the inflationary universe model [12]. This inflation may be the first of the two inflations. The second inflation may involve the direct transition of the superstring from $q_7$ to the dimension four without going through the first two extra dimensions ($B_6 = \pi_{1/2}$, $F_6 = e$, $B_5 = A$, and $F_5 = \nu$). It involves the symmetry mixing between dimensions five and seven, so the first two dimensions ($\pi_{1/2}$, e, $\nu$, and A) are merged with the dimension seven, and become parts of force fields and decay modes for the fermions (heavy leptons and baryons) of $F_7$ with seven auxiliary dimensional orbits. Furthermore, during this second inflation, integer charge leptons, fractional charge quarks, the strong force field, and electroweak force field are generated.

## 3. The periodic table of elementary particles

Leptons and quarks are complementary to each other, expressing different aspects of superstring. The model for leptons and quarks is shown in Fig. 4. The periodic table for elementary particles is shown in Table 2.

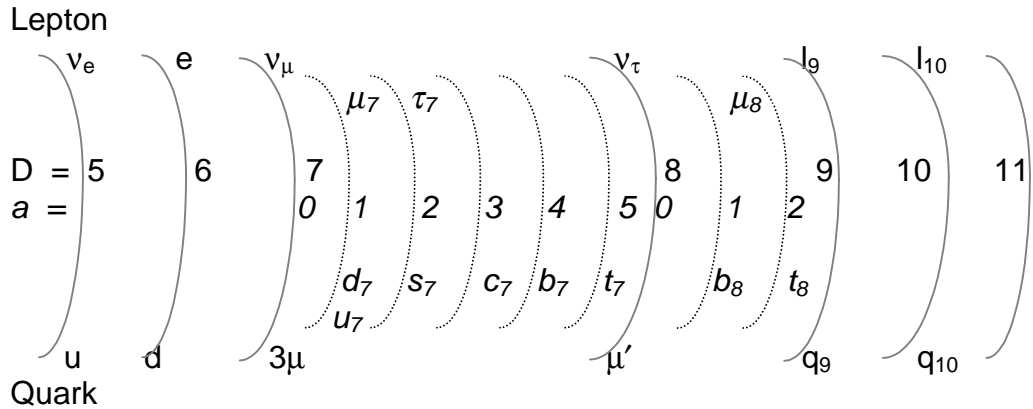

Fig. 4. Leptons and quarks in the dimensional orbits
D = dimensional number, a = auxiliary dimensional number

**Table 2.** The Periodic Table of Elementary Particles
D = dimensional number, a = auxiliary dimensional number

| D | a = 0 | 1 | 2 | a = 0 | 1 | 2 | 3 | 4 | 5 | |
|---|---|---|---|---|---|---|---|---|---|---|
|  | Lepton |  |  | Quark |  |  |  |  |  | Boson |
| 5 | $l_5 = \nu_e$ |  |  | $q_5 = u = 3\nu_e$ |  |  |  |  |  | $B_5 = A$ |
| 6 | $l_6 = e$ |  |  | $q_6 = d = 3e$ |  |  |  |  |  | $B_6 = \pi_{1/2}$ |
| 7 | $l_7 = \nu_\mu$ | $\mu_7$ | $\tau_7$ | $q_7 = 3\mu$ | $u_7/d_7$ $s_7$ | $c_7$ | $b_7$ | $t_7$ |  | $B_7 = Z_L^0$ |
| 8 | $l_8 = \nu_\tau$ | $\mu_8$ |  | $q_8 = \mu'$ | $b_8$ $t_8$ |  |  |  |  | $B_8 = X_R$ |
| 9 | $l_9$ |  |  | $q_9$ |  |  |  |  |  | $B_9 = X_L$ |
| 10 |  |  |  |  |  |  |  |  |  | $B_{10} = Z_R^0$ |
| 11 |  |  |  |  |  |  |  |  |  | $B_{11}$ |

D is the dimensional orbital number for the seven extra space dimensions. The auxiliary dimensional orbital number, a, is for the seven extra auxiliary space dimensions. All gauge bosons, leptons, and quarks are located on the seven dimensional orbits and seven auxiliary orbits. Most leptons are dimensional fermions, while all quarks are the sums of subquarks from the loops around the tori from the compacified extra dimensions. $\nu_e$, e, $\nu_\mu$, and $\nu_\tau$ are dimensional fermions for dimension 5, 6, 7, and 8, respectively. All neutrinos have zero mass because of chiral symmetry.

The dimensional fermions for D = 5 and 6 are neutrino ($l_5$) and electron ($l_6$). Other fermions are generated, so there are more than one fermion in the same dimension. These extra fermions include quarks and heavy leptons ($\mu$ and $\tau$). To generate a quark whose mass is higher than the lepton in the same dimension is to add the lepton to the boson from the combined lepton-antilepton, so the mass of the quark is three times of the mass of the corresponding lepton in the same dimension. The equation for the quark mass is

$$M_{q_D} = 3 M_{l_D} \qquad (9)$$

The mixing between the fifth and the seventh dimensional orbits which is same as the symmetry mixing in the standard model allows the seventh dimensional orbit to be the starting dimensional orbit for the additional orbits, auxiliary orbits with quantum number "a." (The total number of the auxiliary orbits is also seven as the seven dimensional orbits.)

A heavy lepton ($\mu$ or $\tau$) is the combination of the dimensional leptons and the auxiliary dimensional leptons. In the same way, a heavy quark is the combination of the dimensional quarks from Eq.(9) and the auxiliary quarks. The mass of the auxiliary dimensional fermion (AF for both heavy lepton and heavy quark) is generated from the corresponding dimensional boson as follows.

$$M_{AF_{D,a}} = \frac{M_{B_{D-1,0}}}{\alpha_a} \sum_{a=0}^{a} a^4 \quad , \tag{10}$$

where $\alpha_a$ = auxiliary dimensional fine structure constant, and a = auxiliary dimension number = 0 or integer. The first term, $\frac{M_{B_{D-1,0}}}{\alpha_a}$, of the mass formula (Eq.(10)) for the auxiliary dimensional fermions is derived from the mass equation, Eq. (1), for the dimensional fermions and bosons. The second term, $\sum_{a=0}^{a} a^4$, of the mass formula is for Bohr-Sommerfeld quantization for a charge - dipole interaction in a circular orbit as described by A. Barut [13]. $1/\alpha_a$ is 3/2. The coefficient, 3/2, is to convert the dimensional boson mass to the mass of the auxiliary dimensional fermion in the higher dimension by adding the boson mass to its fermion mass which is one-half of the boson mass. The formula for the mass of auxiliary dimensional fermions (AF) becomes

$$\begin{aligned} M_{AF_{D,a}} &= \frac{M_{B_{D-1,0}}}{\alpha_a} \sum_{a=0}^{a} a^4 \\ &= \frac{3}{2} M_{B_{D-1,0}} \sum_{a=0}^{a} a^4 \\ &= \frac{3}{2} M_{F_{D,0}} \alpha_D \sum_{a=0}^{a} a^4 \end{aligned} \tag{11}$$

When the mass of this auxiliary dimensional fermion is added to the sum of masses from the corresponding dimensional fermions (zero auxiliary dimension number) with the same electric charge and the same dimension, the fermion mass formula for heavy leptons and quarks is derived as follows.

$$\begin{aligned} M_{F_{D,a}} &= \sum M_{F_{D,0}} + M_{AF_{D,a}} \\ &= \sum M_{F_{D,0}} + \frac{3}{2} M_{B_{D-1,0}} \sum_{a=0}^{a} a^4 \\ &= \sum M_{F_{D,0}} + \frac{3}{2} M_{F_{D,0}} \alpha_D \sum_{a=0}^{a} a^4 \end{aligned} \quad (12)$$

Each fermion can be defined by dimensional numbers (D's) and auxiliary dimensional numbers (a's). Heavy leptons and quarks consist of one or more D's and a's. The compositions and calculated masses of leptons and quarks are listed in Table 3.

**Table 3.** The Compositions and the Constituent Masses of Leptons and Quarks

D = dimensional number and a = auxiliary dimensional number

|  | $D_a$ | Composition | Calc. Mass |
|---|---|---|---|
| Leptons | $D_a$ for leptons |  |  |
| $\nu_e$ | $5_0$ | $\nu_e$ | 0 |
| e | $6_0$ | e | 0.51 MeV (given) |
| $\nu_\mu$ | $7_0$ | $\nu_\mu$ | 0 |
| $\nu_\tau$ | $8_0$ | $\nu_\tau$ | 0 |
| $\mu$ | $6_0 + 7_0 + 7_1$ | $e + \nu_\mu + \mu_7$ | 105.6 MeV |
| $\tau$ | $6_0 + 7_0 + 7_2$ | $e + \nu_\mu + \tau_7$ | 1786 MeV |
| $\mu'$ | $6_0 + 7_0 + 7_2 + 8_0 + 8_1$ | $e + \nu_\mu + \mu_7 + \nu_\tau + \mu_8$ | 136.9 GeV |
| Quarks | $D_a$ for quarks |  |  |
| u | $5_0 + 7_0 + 7_1$ | $u_5 + q_7 + u_7$ | 330.8 MeV |
| d | $6_0 + 7_0 + 7_1$ | $d_6 + q_7 + d_7$ | 332.3 MeV |
| s | $6_0 + 7_0 + 7_2$ | $d_6 + q_7 + s_7$ | 558 MeV |
| c | $5_0 + 7_0 + 7_3$ | $u_5 + q_7 + c_7$ | 1701 MeV |
| b | $6_0 + 7_0 + 7_4$ | $d_6 + q_7 + b_7$ | 5318 MeV |
| t | $5_0 + 7_0 + 7_5 + 8_0 + 8_2$ | $u_5 + q_7 + t_7 + q_8 + t_8$ | 176.5 GeV |

The lepton for dimension five is $\nu_e$, and the quark for the same dimension is $u_5$, whose mass is equal to 3 $M\nu_e$ from Eq. (9). The lepton for the dimension six is e, and the quark for this dimension is $d_6$. $u_5$ and $d_6$ represent the "light quarks" or "current quarks" which have low masses. The dimensional lepton for the dimensions seven is $\nu_\mu$. All $\nu$'s become massless by the chiral symmetry to preserve chirality. The auxiliary dimensional leptons (Al) in the dimension seven are $\mu_7$ and $\tau_7$ whose masses can be calculated by Eq. (13) derived from Eq. (11).

$$M_{Al_{7,a}} = \frac{3}{2} M_{B_{6,0}} \sum_{a=0}^{a} a^4$$
$$= \frac{3}{2} M_{\pi_{1/2}} \sum_{a=0}^{a} a^4 ,  \quad (13)$$

where a = 1, 2 for $\mu_7$ and $\tau_7$, respectively. The mass of $\mu$ is the sum of e and $\mu_7$, and the mass of $\tau$ is the sum of e + $\tau_7$, as in Eq.(12).

For heavy quarks, $q_7$ (the dimensional fermion, $F_7$, for quarks in the dimension seven) is 3$\mu$ instead of massless 3$\nu$ as in Eq. (9). According to the mass formula, Eq. (11), of the auxiliary fermion, the mass formula for the auxiliary quarks, $Aq_{7,a}$, is as follows.

$$M_{Aq_{7,a}} = \frac{3}{2} M_{q_7} \alpha_7 \sum_{a=0}^{a} a^4$$
$$= \frac{3}{2} (3 M_\mu) \alpha_w \sum_{a=0}^{a} a^4 ,  \quad (14)$$

where $\alpha_7 = \alpha_w$, and a = 1, 2, 3, 4, and 5 for $u_7/d_7$, $s_7$, $c_7$, $b_7$, and $t_7$, respectively.

The dimensional lepton for the dimension eight is $\nu_\tau$, whose mass is zero to preserve chirality. The heavy lepton for the dimensional eight is $\mu'$ as the sum of e, $\mu$, and $\mu_8$ (auxiliary dimensional lepton). Because there are only three families for leptons, $\mu'$ is the extra lepton, which is "hidden". $\mu'$ can appear only as $\mu$ + photon. The pairing of $\mu$ + $\mu$ from the hidden $\mu'$ and regular $\mu$ may account for the occurrence of same sign dilepton in the high energy level [14]. For the dimension eight, $q_8$ (the $F_8$ for quarks) is $\mu'$ instead of $3\mu'$, because the hiding of $\mu'$ allows $q_8$ to be $\mu'$. The hiding of $\mu'$ for leptons is balanced by the hiding of $b_8$ for quarks.

The calculated masses are in good agreement with the observed constituent masses of leptons and quarks [2,15]. The mass of the top quark found by Collider Detector Facility is 176 ± 13 GeV [2] in a good agreement with the calculated value, 176.5 GeV.

## *4.    Hadrons*

As molecules are the composites of atoms, hadrons are the composites of elementary particles. Hadron can be represented by elementary particles in many different ways. One way to represent hadron is through the nonrelativistic constituent quark model where the mass of a hadron is the sum of the masses of quarks plus a relatively small binding energy. $\pi_{1/2}$ (u or d pseudoscalar quark), u, d, s, c, b, and t quarks mentioned above are nonrelativistic constituent quarks. On the other hand, except proton and neutron, all hadrons are unstable, and decay eventually into low-mass quarks and leptons. Is a hadron represented by all nonrelativistic quarks or by low-mass quarks and leptons? This paper proposes a dual formula: the full quark formula for all nonrelativistic constituent quarks ($\pi_{1/2}$, u, d, s, c, and b) and the basic fermion formula for the lowest- mass quark ($\pi_{1/2}$ and u) and lepton (e). The calculation of the masses of hadrons requires both formulas. The full quark formula sets the initial mass for a hadron. This initial mass is matched by the mass resulted from the combination of various particles in the basic fermion formula.  The mass of a hadron is the mass plus the binding energy in the basic fermion formula.

$$M_{\text{full quark formula}} \approx M_{\text{basic fermion formula}}$$

$$M_{\text{hadron}} = M_{\text{basic fermion formula}} + \text{binding energy} \quad (15)$$

The full quark formula consists of the vector quarks (u, d, s, c, and b) and pseudoscalar quark, $\pi_{1/2}$ (70.03 MeV), which is $B_6$ (dimension boson in the dimension six).  The strong interaction converts $B_6$ ($\pi_{1/2}$) into pseudoscalar u and d quark in pseudoscalar $\pi$ meson.  The combination of the pseudoscalar quark ($\pi_{1/2}$) and vector quarks (q) results in hybrid quarks (q') whose mass is the average mass of pseudoscalar quark and vector quark.

$$Mq' = 1/2 \, ( Mq + M\pi_{1/2}) \quad (16)$$

Hybrid quarks include u', d', and s' whose masses are 200.398, 201.164, and 314.148 MeV, respectively.  For baryons other than n and p, the full quark formula is the combination of vector quarks, hybrid quarks (u' and d'), and pseudoscalar quark ($\pi_{1/2}$).  For example, $\Lambda^\circ$ (uds) is u'd's3, where 3 denotes 3 $\pi_{1/2}$.  For pseudoscalar mesons (J = 0), the full quark formula is the combination of $\pi_{1/2}$ and q' (u', d' and s') or $\pi_{1/2}$ alone.  For vector mesons (J > 0), the full quark formula is the combination of vector quarks (u, d, s, c, and b) and $\pi_{1/2}$.  For examples, $\pi^\circ$ is 2, $\eta$ (1295, J =0) is u'u'd'd'8, and $K_1$ (1400, J=1) is ds8.  The compositions of hadrons from the particles of the full quark formula are listed in Table 4.

**Table 4.**  Particles for the full quark formula

|  | $\pi_{1/2}$ | u', d' | s' | u, d | s | c, b |
|---|---|---|---|---|---|---|
| mass (MeV) | 70.025 | 200.40, 201.16 | 314.15 | 330.8, 332.3 | 558 | 1701, 5318 |
| n and p |  |  |  | √ |  |  |
| baryons other than n and p |  | √ | √ |  | √ | √ |
| mesons (J = 0) except $c\bar{c}$ and $b\bar{b}$ |  | √ | √ | √ |  | √ |
| mesons (J > 0) and $c\bar{c}$ and $b\bar{b}$ |  | √ |  | √ | √ | √ |

Different energy states in hadron spectroscopy are the results of differences in the numbers of $\pi_{1/2}$. The higher the energy state, the higher the number of $\pi_{1/2}$. The number of $\pi_{1/2}$ attached to q or q' in the full quark formula is restricted. The number of $\pi_{1/2}$ can be 0, ±1, ±3, ±4, ±7, and ±3n (n>2). This is the "$\pi_{1/2}$ series." The number, 3, is indicative of a baryon-like (3 quarks in a baryon) number. The number, 4, is the combination of 1 and 3, while the number, 7, is the combination of 3 and 4. Since $\pi_{1/2}$ is essentially pseudoscalar u and d quarks, the $\pi_{1/2}$ series is closely related to u and d quarks. Since s is close to u and d, the $\pi_{1/2}$ series is also related to s quark. . Each presence of u, d, and s associates with one single $\pi_{1/2}$ series. The single $\pi_{1/2}$ series associates with baryons and the mesons with $u\bar{u}$ d $\bar{d}$ s $\bar{s}$, s $\bar{s}$, d $\bar{c}$, u $\bar{c}$, d $\bar{b}$, u $\bar{b}$, s $\bar{c}$, and s $\bar{b}$. The combination of two single $\pi_{1/2}$ series is the double $\pi_{1/2}$ series (0, ±2, ±6,

±8, ±14, and ±6n) associating with u $\bar{d}$ and d $\bar{s}$ mesons. For the mesons (c $\bar{c}$ and b $\bar{b}$) without u, d, or s, the numbers of $\pi_{1/2}$ attached to quarks can be from the single $\pi_{1/2}$ series or the double $\pi_{1/2}$ series.

The basic fermion formula is similar to M. H. MacGregor's light quark model [17], whose calculated masses and the predicted properties of hadrons are in very good agreement with observations. In the light quark model, the mass building blocks are the "spinor" (S with mass 330.4 MeV) and the mass quantum (mass = 70MeV). S has 1/2 spin and neutral charge, while the basic quantum has zero spin and neutral charge. For the basic fermion formula, S is u, the quark with the lowest mass (330.77 MeV), and the basic quantum is $\pi_{1/2}$ (mass = 70.025 MeV). In the basic fermion formula, u and $\pi_{1/2}$ have the same spins and charge as S and the basic quantum, respectively. For examples, in the basic fermion formula, neutron is SSS, and $K_1$ (1400, J=1) is $S_2$ 11, where $S_2$ denotes two S, and 11 denotes 11 $\pi_{1/2}$'s.

In additional to S and $\pi_{1/2}$, the basic fermion formula includes P (positive charge) and N (neutral charge) with the masses of proton and neutron. As in the light quark model, the mass associated with positive or negative charge is the electromagnetic mass, 4.599 MeV, which is nine times the mass of electron. This mass (nine times the mass of electron) is derived from the baryon-like electron that represents three quarks in a baryon and three electron in $d_6$ quark as in Table 2. This electromagnetic mass is observed in the mass difference between $\pi°$ (2) and $\pi^+$ (2+) where + denotes positive charge. The calculated mass different is one electromagnetic mass, 4.599 MeV, in good agreement with the observed mass difference, 4.594 MeV, between $\pi°$ and $\pi^+$. (The values for observed masses are taken from "Particle Physics Summary "[18].) The particles in the basic fermion formula are listed in Table 5.

**Table 5.** Particles in the basic fermion formula

|  | S | $\pi_{1/2}$ | N | P | electromagnetic mass |
|---|---|---|---|---|---|
| mass (MeV) | 330.77 | 70.0254 | 939.565 | 938.272 | 4.599 |

Hadrons are the composites of quarks as molecules composing of atoms. As atoms are bounded together by chemical bonds, hadrons are bounded by "hadronic bonds," connecting the particles in the basic fermion formula. These hadronic bonds are similar to the hadronic bonds in the light quark model.

The hadronic bonds are the overlappings of the auxiliary dimensional orbits. As in Eq (11), the energy derived from the auxiliary orbit for S (u quark) is

$$E_a = (3/2)(3 M_\mu) \alpha_w$$
$$= 14.122 \text{ MeV} \qquad (17)$$

The auxiliary orbit is a charge - dipole interaction in a circular orbit as described by A. Barut [13], so a fermion for the circular orbit and an electron for the charge are embedded in this hadronic bond. The fermion for the circular orbit is the supersymmetrical fermion for the auxiliary dimensional orbit according to Eq (2).

$$M_f = E_a \alpha_w \tag{18}$$

The binding energy (negative energy) for the bond ($S_\daleth S$) between two S's is twice of 14.122 MeV minus the masses of the supersymmetrical fermion and electron.

$$E_{S-S} = -2(E_a - M_f - M_e)$$

$$= -26.384 \text{ MeV} \tag{19}$$

It is similar to the binding energy (-26 MeV) in the light quark model. An example of $S_\daleth S$ bond is in neutron (S – S – S) which has two S – S bonds. The mass of neutron can be calculated as follows.

$$M_n = 3M_S + 2E_{S-S}$$

$$= 939.54 \text{ MeV}, \tag{20}$$

which is in excellent agreement with the observed mass, 939.57 MeV. The mass of proton is the mass of neutron minus the mass difference (three times of electron mass = $M_{3e}$) between u and d quark as shown in Table 2. Proton is represented as S – S – (S -3e). The calculation of the mass of proton is as follows.

$$E_a \text{ for (S-3e)} = (3/2)(3(M_\mu - M_{3e}))\alpha_w$$

$$M_f = E_a \alpha_w$$

$$M_p = 2M_S + M_{(S-3e)} + E_{S_\daleth S} + E_{S_\daleth(S-3e)}$$

$$= 938.21 \text{ MeV} \tag{21}$$

The calculated mass is in a good agreement with the observed mass, 938.27 MeV.

The binding energy for $\pi_{1/2} - \pi_{1/2}$ bond can be derived in the same way as Eqs (17), (18), and (19).

$$E_a = (3/2) M\pi_{1/2} \alpha_w$$

$$M_f = E_a \alpha_w$$

$$E_{\pi_{1/2} - \pi_{1/2}} = -2 (E_a - M_f - M_e)$$

$$= -5.0387 \text{ MeV} \tag{22}$$

It is similar to the binding energy (-5 MeV) in the light quark model. An example for the binding energy of $\pi_{1/2} - \pi_{1/2}$ bond is in $\pi°$. The mass of $\pi°$ can be calculated as follows.

$$M_{\pi°} = 2 M\pi_{1/2} + E_{\pi_{1/2} - \pi_{1/2}}$$

$$= 135.01 \text{ MeV}. \tag{23}$$

The calculated mass of $\pi°$ is in excellent agreement with the observed value, 134.98 MeV. There is one $\pi_{1/2} - \pi_{1/2}$ bond per pair of $\pi_{1/2}$'s, so there are two $\pi_{1/2} - \pi_{1/2}$ bonds for 4 $\pi_{1/2}$'s, and three $\pi_{1/2} - \pi_{1/2}$ bonds for 6 $\pi_{1/2}$'s.

Another bond is $N - \pi_{1/2}$ or $P - \pi_{1/2}$, the bond between neutron or proton and $\pi_{1/2}$. Since N is SSS, $N - \pi_{1/2}$ bond is derived from $S - \pi_{1/2}$. The binding energy of $S - \pi_{1/2}$ is the average between $S-S$ and $\pi_{1/2} - \pi_{1/2}$.

$$E_{S - \pi_{1/2}} = 1/2 ( E_{S-S} + E_{\pi_{1/2} - \pi_{1/2}}) \tag{24}$$

An additional dipole ($e^+ e^-$) is needed to connected $S - \pi_{1/2}$ to neutron.

$$E_{N - \pi_{1/2}} = E_{S - \pi_{1/2}} + 2 M_e$$

$$= -14.689 \text{ MeV}. \tag{25}$$

It is similar to -15 MeV in the light quark model. An example for $P - \pi_{1/2}$ is $\Sigma^+$ which is represented by P4 whose structure is $2 - P - 2$. The 4 $\pi_{1/2}$'s are connected to P with two $P - \pi_{1/2}$ bonds. The mass of $\Sigma^+$ is as follows.

$$M_{\Sigma^+} = M_P + 4 M_{\pi_{1/2}} + 2 E_{N-\pi_{1/2}}$$

$$= 1189.0 \text{ MeV}. \tag{26}$$

The calculated mass is in a good agreement with the observed mass, $1189.4 \pm 0.07$ MeV.

There is N – N hadronic bond between two N's. N has the structure of S – S – S. N – N has a hexagonal structure shown in Fig 5.

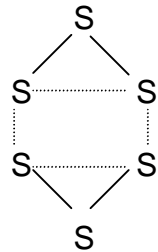

**Fig 5.** The structure of N - N

There are two additional S – S for each N. The total number of S – S bonds between two N's is 4. An example is $\Omega^\circ_c$ (ssc) which has the basic fermion formula of $N_2 S9$ with the structure of $3 - N - N - 6 S$. The mass of $\Omega^\circ_c$ can be calculated as follows.

$$M = 2M_N + M_S + 9M_{\pi_{1/2}} + 4 E_{S-S} + 2 E_{N-\pi_{1/2}}$$

$$= 2705.3 \text{ MeV} \tag{27}$$

The calculated mass for $\Omega^{\circ}_c$ is in a good agreement with the observed mass (2704 ± 4 MeV). For baryons other than p and n, there are two or three N - $\pi_{1/2}$ or P - $\pi_{1/2}$ bonds per baryon. It is used to distinguish J = ½ and J ≥ 3/2. For J = ½ which have asymmetrical spins (two up and one down), there are two bonds for three S in a baryon to represent the asymmetrical spins. For J ≥ 3/2, there are three bonds for three S in a baryon except if there is only one $\pi_{1/2}$, only two bonds exists for a baryon.

Among the particles in the basic fermion formula, there are hadronic bonds, but not all particles have hadronic bonds. In the basic fermion formula, hadronic bonds appear only among the particles that relate to the particles in the corresponding full quark formula. The related particles are the "core" particles that have hadronic bonds, and the unrelated particles are "filler" particles that have no hadronic bonds. In the basic fermion formula, for baryons other than p and n, the core particles are P, N, and $\pi_{1/2}$. For the mesons consisting of u and d quarks, the core particles are $\pi_{1/2}$, S, and N. For the mesons containing one u, d, or s along with s, c, or b, the core particles are S and N, and no hadronic bond exist among $\pi_{1/2}$'s , which are the filler particles. For the mesons (c $\bar{c}$ and b $\bar{b}$), the only hadronic bond is N – N. The occurrences of hadronic bonds are listed in Table 6.

**Table 6.** Hadronic bonds in hadrons

|  | S–S | $\pi_{1/2} - \pi_{1/2}$ | N(P) - $\pi_{1/2}$ | N–N |
|---|---|---|---|---|
| binding energy (MeV) | -26.384 | -5.0387 | -14.6894 | 2 S–S per N |
| baryons other than n and p | √ |  | √ | √ |
| mesons with u and d only | √ | √ |  | √ |
| mesons containing one u, d, or s along with s, c, or b | √ |  |  | √ |
| $c \bar{c}$ or $b \bar{b}$ mesons |  |  |  | √ |

An example is the difference between $\pi$ and $f_0$. The decay modes of $f_0$ include the mesons of s quarks from K meson. Consequently, there is no $\pi_{1/2}$-$\pi_{1/2}$ for $f_0$. The basic fermion formula for $f_0$ is 14. The mass of $f_0$ is as follows.

$$M = 14 M\pi_{1/2}$$
$$= 980.4 \text{ MeV} \qquad (28)$$

The observed mass is $980 \pm 10$ MeV.

In additional to the binding energies for hadronic bonds, hadrons have Coulomb energy (-1.2 MeV) between positive and negative charges and magnetic binding energy (±2MeV per interaction) for S – S from the light quark model [17]. In the light quark model, the dipole moment of a hadron can be calculated from the magnetic binding energy. Since in the basic fermion formula, magnetic binding energy becomes a part of hadronic binding energy as shown in Eq (19), magnetic binding energy for other baryons is the difference in magnetic binding energy between a baryon and n or p. If a baryon has a similar dipole moment as p or n, there is no magnetic binding energy for the baryon. An example for Coulomb energy and magnetic binding energy is $\Lambda$ (uds, J=1/2) whose formula is P3- with the structure of 2 – P – 1- where "-" denotes negative charge. The dipole moment of $\Lambda$ is –6.13 $\mu_N$, while the dipole moment of proton (P) is 2.79 $\mu_N$ [18]. According to the light quark model, this difference in dipole moment represents -6 MeV magnetic binding energy. The Coulomb energy between the positive charge P and the negative charge 1- is –1.2 MeV. The electromagnetic mass for 1- is 4.599 MeV. The mass of $\Lambda$ is calculated as follows.

$$M_\Lambda = M_P + 3M\pi_{1/2} + M_{e.m.} + 2 E_{N-\pi_{1/2}} + E_{mag} + E_{coul}$$

$$= 1116.4 \text{ MeV} \tag{29}$$

The observed mass is 1115.7 $\pm$ 0.0006 MeV.

An example of the dual representation of the full quark formula and the basic fermion formula as expressed by Eq. (15) is $\Lambda$ (uds). The full quark formula for $\Lambda$ is u'd's'3 with mass of 1169.9 MeV. This mass is matched by the mass of the basic fermion formula, P3- with the mass of 1152.9 MeV. The final mass of $\Lambda$ is 1152.9 MeV plus the various binding energies.

Table 7 is the results of calculation for the masses of baryons selected from Ref. (18). The hadrons selected are the hadrons with precise observed masses and known quantum states such as isospin and spin.

**Table 7.** The masses of baryons

| Baryons | I($J^P$) | Full quark formula | Basic fermion formula | Calculated mass | Observed mass | Difference |
|---|---|---|---|---|---|---|
| n and p | | | | | | |
| n | 1/2(½$^+$) | udd | SSS | 939.54 | 939.57 | -0.03 |
| p | ½(½$^+$) | uud | SSS-3e | 938.21 | 938.27 | -0.06 |
| uds, uus, dds | | | | | | |
| $\Lambda°$ | 0(½$^+$) | u'd's3 | P3- | 1116.4 | 1115.7 | 0.7 |
| $\Sigma^+$ | 1(½$^+$) | u'u's4 | P4 | 1189.0 | 1189.4 | -0.4 |
| $\Sigma°$ | 1(½$^+$) | u'd's4 | P4- | 1192.4 | 1192.6 | -0.2 |
| $\Sigma^-$ | 1(½$^+$) | d'd's4 | N4- | 1194.9 | 1197.4 | -2.5 |
| $\Sigma^+$ | 1(3/2$^+$) | u'u's7 | P7 | 1384.4 | 1382.8 | 1.6 |
| $\Sigma°$ | 1(3/2$^+$) | u'd's7 | P7- | 1381.8 | 1383.7 | -1.9 |
| $\Sigma^-$ | 1(3/2$^+$) | d'd's7 | N7- | 1390.3 | 1387.2 | 3.1 |
| $\Lambda°$ ($s_{01}$) | 0(½-) | u'd's7 | P7- | 1402.5 | 1407.0 | -4.5 |
| $\Lambda°$ ($D_{03}$) | 0(3/2$^-$) | u'd's9 | N9 | 1525.7 | 1519.5 | 6.2 |
| uss, dss | | | | | | |
| $\Xi°$ | ½(½$^+$) | u'ss | NS1 | 1311.0 | 1314.9 | -3.9 |
| $\Xi^-$ | ½(½$^+$) | d'ss | NS1- | 1315.6 | 1321.3 | -5.7 |
| $\Xi°$ | ½(3/2$^+$) | u'ss4 | N9+- | 1533.7 | 1531.8 | 1.9 |
| $\Xi^-$ | ½(3/2$^+$) | d'ss4 | N10- | 1530.3 | 1535.0 | -4.7 |
| $\Xi°$ | ½(3/2$^-$) | u'ss9 | $N_2$1+- | 1822.2 | 1823.0 | -0.8 |
| udc, ddc, uuc | | | | | | |
| $\Lambda^+_c$ | 0(½$^+$) | u'd'c3 | PS15 | 2290.0 | 2284.9 | 5.1 |
| $\Sigma_c^0$ | 1(½$^+$) | d'd'c7 | $N_2$10 | 2452.5 | 2452.1 | 0.4 |
| $\Sigma_c^{++}$ | 1(½$^+$) | u'u'c7 | $N_2$10++ | 2452.5 | 2452.9 | -0.4 |
| $\Sigma_c^+$ | 1(½$^-$) | u'd'c7 | $N_2$10+ | 2449.1 | 2453.5 | -4.4 |
| $\Lambda^+_c$ | 0(½$^-$) | u'd'c9 | $N_2$12+ | 2589.1 | 2593.6 | -4.5 |
| usc, dsc | | | | | | |
| $\Xi^+_c$ | ½(½$^+$) | u'sc1 | $NS_3$8+ | 2467.3 | 2465.6 | 1.7 |
| $\Xi°_c$ | ½(½$^+$) | d'sc1 | N $S_3$8+- | 2470.7 | 2470.3 | 0.4 |
| sss | | | | | | |
| $\Omega^-$ | 0(3/2$^+$) | sss | NS6- | 1665.7 | 1672.5 | -6.8 |
| ssc | | | | | | |
| $\Omega°_c$ | 0(½$^+$) | ssc | $N_2$S9 | 2705.2 | 2704.0 | 1.2 |
| udb | | | | | | |
| $\Lambda°_b$ | 0(½$^+$) | u'd'b1 | $N_2$S $_9$13+- | 5639.5 | 5641.0 | -1.5 |

The full quark formula for baryons other than p and n involves hybrid quarks (u' and d'), vector quarks (s, c, and b), and pseudoscalar quark ($\pi_{1/2}$). The whole spectrum of baryons (other than p and n) follows the single $\pi_{1/2}$ series with 0, 1, 3, 4, 7, and 3n. $\pi_{1/2}$ is closely related to u and d, so the lowest energy state baryons ($\Lambda$ and $\Sigma$) that contain low mass quark and high number of u or d have high number of $\pi_{1/2}$. . On the contrary, the lowest energy state baryons ($\Omega$, $\Omega_c$, and $\Lambda_b$) that contain high mass quarks and no or low number of u and d have low number of $\pi_{1/2}$ or no $\pi_{1/2}$. $\pi_{1/2}$'s are added to the lowest energy state baryons to form the high-energy state baryons.

The basic fermion formula includes P, N, S, and $\pi_{1/2}$. If the dominant decay mode includes p, the basic fermion formula includes P. If the dominant decay mode includes n, the basic fermion formula includes N. If the decay mode does not include either n or p directly, the baryon with the dipole moment similar to p or n has the basic fermion formula with P or N, respectively. If the dipole moment is not known, the designation of P or N for the higher energy state baryon follows the designation of P or N for the corresponding lower energy state baryon. Only N is used in the multiple nucleons such as $N_2$ (two N's).

The result of calculated masses for light unflavored mesons is listed in Table 8. Light unflavored mesons have zero strange, charm, and bottom numbers.

**Table 8.** Light unflavored mesons

| Meson | I ($J^{pc}$) | Full quark formula | Basic fermion formula | Calculated mass. | Observed mass | Difference |
|---|---|---|---|---|---|---|
| J =0, only u, d, or lepton in decay mode | | | | | | |
| $\pi^°$ | 1 ($0^{-+}$) | 2 | 2 | 135.01 | 134.98 | 0.04 |
| $\pi^\pm$ | 1 ($0^-$) | 2 | 2± | 139.61 | 139.57 | 0.04 |
| η | 0 ($0^{-+}$) | 8 | 8+- | 548.0 | 547.5 | 0.5 |
| η' | 0 ($0^{-+}$) | 14 | 14+- | 958.1 | 957.8 | 0.3 |
| η | 0 ($0^{-+}$) | u'u'u'd'd'd'2 | $S_3$ 5+- | 1292.6 | 1295.0 | -2.4 |
| $f_0$ | 0 ($0^{++}$) | u'u'u'd'd'd'6 | $S_4$ 4 | 1514.0 | 1503.0 | 11.0 |
| J >0, only u, d, or lepton in decay modes | | | | | | |
| ρ | 1 ($1^-$) | ud2 | $S_2$ 2 | 770.2 | 768.5 | 1.7 |
| ω | 0 ($1^-$) | ud2 | $S_2$ 2+- | 785.4 | 781.9 | 3.5 |
| $h_1$ | 0 ($1^{+-}$) | ud8 | $S_2$ 8 | 1176.2 | 1170.0 | 6.2 |
| ω | 0 ($1^-$) | uudd2 | $S_3$ 7+- | 1415.6 | 1419.0 | -3.4 |
| ω | 0 ($1^-$) | uudd6 | $S_4$ 6 | 1650.0 | 1649.0 | 1.0 |
| $ω_3$ | 0 ($3^-$) | uudd6 | $S_4$ 6 | 1662.0 | 1667.0 | -5.0 |
| J = 0, u and d in major decay modes, s in minor decay modes | | | | | | |
| $f_0$ | 0 ($0^{++}$) | u'd's'4 | 14 | 980.4 | 980.0 | 0.4 |
| $a_0$ | 1 ($0^{++}$) | u'd's'4 | 14 | 985.4 | 983.5 | 1.9 |
| η | 0 ($0^{-+}$) | u'u'd'd's's' | $S_2$ 11+- | 1418.5 | 1415.0 | 3.5 |
| J >o, u and d in major decay modes, s in minor decay mode | | | | | | |
| $a_1$ | 1 ($1^{++}$) | uds | S13 | 1232.7 | 1230.0 | 2.7 |
| $b_1$ | 1 ($1^{+-}$) | uds | S13 | 1232.7 | 1231.0 | 1.7 |
| $f_2$ | 0 ($2^{++}$) | uds1 | $S_2$ 9 | 1278.4 | 1275.0 | 3.4 |
| $f_1$ | 0 ($1^{++}$) | uds1 | $S_2$ 9+- | 1278.4 | 1282.2 | -3.8 |
| $a_2$ | 1 ($2^{++}$) | uds1 | S14 | 1316.2 | 1318.1 | -1.9 |
| $f_1$ | 0 ($1^{++}$) | uds4 | $S_3$ 7 | 1429.7 | 1426.8 | 2.9 |
| ρ | 1 ($1^-$) | uds4 | $S_2$ 12 | 1475.5 | 1465.0 | 10.5 |
| $\pi_2$ | 1 ($2^{-+}$) | uds7 | $S_2$ 15 | 1685.5 | 1670.0 | 15.5 |
| $ρ_3$ | 1 ($3^-$) | uds7 | $S_2$ 15 | 1693.5 | 1691.0 | 3.5 |
| ρ | 1 ($1^-$) | uds7 | $S_2$ 15+- | 1698.6 | 1700.0 | -1.4 |
| $f_4$ | 0 (4++) | uds12 | $S_2$ 20+- | 2043.7 | 2044.0 | -0.3 |
| s in major decay modes | | | | | | |
| Φ | 0 ($1^-$) | ss-1 | $S_3$ +- | 1017.6 | 1019.4 | -1.8 |
| $f_1$ | 0 ($1^{++}$) | ss7 | $S_4$ 4 | 1524.0 | 1512.0 | 12.0 |
| $f_2$ | 0 ($2^{++}$) | ss7 | $S_4$ 4+- | 1532.0 | 1525.0 | 7.0 |
| Φ | 0 ($1^-$) | ss9 | $S_4$ 6+- | 1672.1 | 1680.0 | -7.9 |
| $Φ_3$ | 0 ($3^-$) | ss12 | $S_6$ | 1852.7 | 1854.0 | -1.3 |
| $f_2$ | 0 ($2^{++}$) | ss15 | $S_6$ 2 | 2000.7 | 2011.0 | -10.3 |
| $f_2$ | 0 ($2^{++}$) | ss18 | N $S_3$ 6 | 2299.3 | 2297.0 | 2.3 |
| $f_2$ | 0 ($2^{++}$) | ss18 | N $S_4$ 2+- | 2331.5 | 2339.0 | -7.5 |

The full quark formulas for light unflavored mesons are different for different decay modes. There are three different types of the full quark formula for light unflavored mesons. Firstly, when the decay mode is all leptons or all mesons with u and d quarks, the full quark formula consists of $\pi_{1/2}$, u', d', u, and d quarks, Secondly, when u, d, and leptons are in the major decay modes, and s is in the minor decay modes, the full quark formula consists of $\pi_{1/2}$, u', d', s', u, d, and s quarks. The most common full quark formula for such mesons is uds, which is essentially ½ (u $\bar{u}$ d $\bar{d}$ s $\bar{s}$). Finally, when s quarks are in the major or all decay mode, the full quark formula consists of $\pi_{1/2}$'s and s quarks. When J = 0, the full quark formula includes $\pi_{1/2}$'s and hybrid quarks (u', d', and s'). When J > 0, the full quark formula includes $\pi_{1/2}$'s and vector quarks (u, d, and s). Since there are double presence of u and d quarks for the full quark formula with u and d quark, it follows the double $\pi_{1/2}$ series: 0, 2, 6, 8, 14, and 6n. The full quark formula with odd presence of u, d, and s follows the single $\pi_{1/2}$ series: 0, ±1, ±3, ±4, ±7, and ±3n.

The basic fermion formula includes $\pi_{1/2}$'s, S, and N. $\pi_{1/2} - \pi_{1/2}$ hadronic bond exists in the full quark formula with u and d quarks, and does not exist in the full quark formula with s quark. S – S bond exists in all formula. Light unflavored mesons decay into symmetrical low mass mesons, such as 2γ, 3$\pi^o$, and $\pi^0$2γ from η, or asymmetrical high and low mass mesons, such as asymmetrical $\pi^+\pi\eta$, $\rho^0\gamma$, and $\pi^0\pi^0\eta$ from η'. To distinguish the asymmetrical decay modes from the symmetrical decay modes, one "counter $\pi_{1/2} - \pi_{1/2}$ hadronic bond" is introduced in η' [17]. The binding energy for the counter $\pi_{1/2} - \pi_{1/2}$ hadronic bond is 5.04 MeV, directly opposite to −5.04 MeV for the $\pi_{1/2} - \pi_{1/2}$ hadronic bond. All light unflavored mesons with asymmetrical decay modes include this counter $\pi_{1/2} - \pi_{1/2}$ hadronic bonds. For an example, the mass of η' (14+-) is calculated as follows.

$$M_{\eta'} = 14 M_{\pi_{1/2}} + 2 M_{e.m.} + E_{e.m.} + 7E_{\pi_{1/2} - \pi_{1/2}} - E_{\pi_{1/2} - \pi_{1/2}}$$

$$= 958.2 \text{ MeV} \qquad (30)$$

The observed mass is 957.8 ± 0.14 MeV.

The result of the mass calculation for the mesons consisting of u, d, or s with s, c, or b is listed in Table 9.

**Table 9.** Mesons with s, c, and b

| Meson | $J^{PC}$ | Full quark formula | Basic fermion formula | Calculated mass. | Observed mass | Difference |
|---|---|---|---|---|---|---|
| Light strange mesons | | | | | | |
| $K^\pm$ | $0^-$ | 7 | $7\pm$ | 494.8 | 493.7 | 1.1 |
| $K^0$ | $0^-$ | 7 | $7+-$ | 498.2 | 497.7 | 0.5 |
| $K^*$ | $0^+$ | d's'14 | $S_3\,7$ | 1429.7 | 1429.0 | 0.7 |
| $K^*$ | $1^-$ | us | $S8\pm$ | 895.6 | 891.6 | 4.0 |
| $K^*$ | $1^-$ | ds | $S8$ | 899.0 | 896.1 | 2.9 |
| $K_1$ | $1^+$ | ds6 | $S_2\,9$ | 1273.4 | 1273.0 | 0.4 |
| $K_1$ | $1^+$ | ds8 | $S_2\,11$ | 1405.4 | 1402.0 | 3.4 |
| $K^*$ | $1^-$ | ds8 | $S_2\,11+-$ | 1413.4 | 1412.0 | 1.4 |
| $K^{*\pm}$ | $2^+$ | us8 | $N7\pm$ | 1434.3 | 1425.4 | 8.9 |
| $K^*$ | $2^+$ | ds8 | $N7+-$ | 1437.7 | 1432.4 | 5.3 |
| $K^*$ | $1^-$ | ddss | $S_3\,11+-$ | 1717.8 | 1714.0 | 3.8 |
| $K_2$ | $2^-$ | ddss | $N12$ | 1779.9 | 1773.0 | 6.9 |
| $K_3^*$ | $3^-$ | ddss | $N12$ | 1779.9 | 1780.0 | -0.1 |
| $K_2$ | $2^-$ | ds14 | $S_4\,8+-$ | 1812.1 | 1816.0 | -3.9 |
| $K_4^*$ | $4^+$ | ds18 | $S_5\,7+-$ | 2046.5 | 2045.0 | 1.5 |
| Charmed mesons | | | | | | |
| $D^0$ | $0^-$ | u'c1 | $S_6\,+-$ | 1860.7 | 1864.5 | -3.8 |
| $D^\pm$ | $0^-$ | d'c1 | $S_6\,\pm$ | 1857.3 | 1869.3 | -12.0 |
| $D^{*0}$ | $1^-$ | uc1 | $S_6\,2+-$ | 2000.7 | 2006.7 | -6.0 |
| $D^{*\pm}$ | $1^-$ | dc1 | $S_6\,2\pm$ | 1997.4 | 2010.0 | -12.6 |
| $D_1$ | $1^+$ | uc7 | $S_6\,8+-$ | 2420.9 | 2422.2 | -1.3 |
| $D_2^*$ | $2^+$ | uc7 | $N\,S_4\,4$ | 2463.6 | 2458.9 | 4.7 |
| $D_2^{*\pm}$ | $2^+$ | dc7 | $N\,S_4\,4\pm$ | 2468.2 | 2459.0 | 9.2 |
| Charmed strange mesons | | | | | | |
| $D_s$ | $0^-$ | s'c1 | $S_5\,6$ | 1968.5 | 1968.5 | 0.0 |
| $D_{s1}$ | $1^+$ | sc4 | $NS18$ | 2535.4 | 2535.4 | 0.0 |
| Bottom mesons | | | | | | |
| $B^\pm$ | $0^-$ | u'b | $N_4\,S20\pm$ | 5283.1 | 5278.9 | 4.2 |
| $B^0$ | $0^-$ | d'b | $N_4\,S\,20$ | 5278.5 | 5279.2 | -0.7 |
| $B^*$ | $1^-$ | db | $N_6$ | 5320.8 | 5324.8 | -4.0 |
| $B_s$ | $0^-$ | s'b | $N_5\,S_3$ | 5373.5 | 5369.3 | 4.2 |

The full quark formula for these mesons contains $\pi_{1/2}$'s, u', d', s', u, d, s, c, and b. The full quark formula for the mesons with J = 0 contains $\pi_{1/2}$, hybrid quarks (u', d', and s'), c, and b quarks. The full quark formula with J > 0 contains $\pi_{1/2}$'s and vector quarks (u, d, s, c, and b). Strange mesons have double presence of d and s, so it follows the double $\pi_{1/2}$ series. All other mesons have single presence of u, d, or s, so they follow the single $\pi_{1/2}$ series. The basic fermion formula consists of $\pi_{1/2}$'s, S, and N. There is no $\pi_{1/2} - \pi_{1/2}$ bond. There are S – S and N – N bonds. For example, the mass of $B_s$ ($N_5 S_3$) is calculated as follows.

$$M = 5M_N + 3M_S + (10 + 2) M_{S-S}$$
$$= 5373.5 \text{ MeV} \qquad (31)$$

There are 10 S – S bonds for $N_5$ (two S – S bonds for each N) and 2 S – S bonds for $S_3$. The observed mass is 5369.3 ± 2 MeV.

The result of mass calculation for $c\bar{c}$ and $b\bar{b}$ mesons is listed in Table 10.

**Table 10.** c $\bar{c}$ and b $\bar{b}$ mesons

| Meson | J $^{pc}$ | Full quark formula | Basic fermion formula | Calculated mass. | Observed mass | Difference |
|---|---|---|---|---|---|---|
| c $\bar{c}$ mesons | | | | | | |
| $\eta_c$ (1s) | $0^{-+}$ | cc-6 | N S$_3$ 15 | 2982.3 | 2979.8 | 2.5 |
| J/$\psi$ | $1^{--}$ | cc-3 | N$_2$ S$_4$ | 3096.7 | 3096.9 | -0.2 |
| $\chi_c$ (1p) | $0^{++}$ | cc2 | N$_2$ S$_2$ 14 | 3415.5 | 3415.1 | 0.4 |
| $\chi_c$ (1p) | $1^{++}$ | cc3 | N$_2$ S$_4$ 6 | 3516.8 | 3510.5 | 6.5 |
| $\chi_{c2}$ (1p) | $2^{++}$ | cc4 | N$_2$ S$_2$16 | 3555.5 | 3556.2 | -0.7 |
| U$_{2s}$ | $1^{--}$ | cc6 | N$_3$ S10 | 3691.4 | 3686.0 | 5.4 |
| $\psi$ | $1^{--}$ | cc7 | N$_3$ S11 | 3761.4 | 3769.9 | -8.5 |
| $\psi$ | $1^{--}$ | cc12 | N$_4$ 7 | 4037.4 | 4040.0 | -2.6 |
| $\psi$ | $1^{--}$ | cc14 | N$_4$ S4 | 4158.1 | 4159.0 | -0.9 |
| $\psi$ | $1^{--}$ | cc18 | N$_4$ S$_2$ 3 | 4418.8 | 4415.0 | 3.8 |
| b $\bar{b}$ mesons | | | | | | |
| $\Upsilon$ (1s) | $1^{--}$ | bb-12 | N$_7$ S $_6$ 18 | 9452.7 | 9460.4 | -7.7 |
| $\chi_b$ (1p) | $0^{++}$ | bb-4 | N$_{10}$ S $_3$ | 9860.3 | 9859.8 | 0.5 |
| $\chi_{b1}$ (1p) | $1^{++}$ | bb-3 | N$_{10}$ S10 | 9899.0 | 9891.9 | 7.1 |
| $\chi_{b2}$ (1p) | $2^{++}$ | bb-3 | N$_{10}$ 15 | 9918.4 | 9913.2 | 5.2 |
| $\Upsilon$ (2s) | $1^{--}$ | bb-1 | N$_{10}$ S$_2$ 7 | 10019.7 | 10023.3 | -3.6 |
| $\chi_{b0}$ (1p) | $0^{++}$ | bb2 | N$_{10}$ S$_2$ 10 | 10229.8 | 10232.1 | -2.3 |
| $\chi_{b1}$ (2p) | $1^{++}$ | bb2 | N$_{10}$ S$_4$ 1 | 10261.1 | 10255.2 | 5.9 |
| $\chi_{b2}$ (2p) | $2^{++}$ | bb2 | N$_{10}$ S$_4$ 1 | 10261.1 | 10268.5 | -7.4 |
| $\Upsilon$ (3s) | $1^{--}$ | bb3 | N$_{10}$ S$_3$ 7 | 10350.5 | 10355.3 | -4.8 |
| $\Upsilon$ (4s) | $1^{--}$ | bb7 | N$_{11}$ S7 | 10575.7 | 10580.0 | -4.3 |
| $\Upsilon$ | $1^{--}$ | bb12 | N$_{12}$ 3 | 10851.6 | 10865.0 | -13.4 |
| $\Upsilon$ | $1^{--}$ | bb14 | N$_{11}$ S$_3$ 4 | 11027.2 | 11019.0 | 8.2 |

The full quark formula contains c or b. Since it contains no u, d, or s related to $\pi_{1/2}$, it follows a mixed $\pi_{1/2}$ series from both the single $\pi_{1/2}$ series and the double $\pi_{1/2}$ series. Since c and b are high mass quarks unlike the low mass u and d quarks that relate to $\pi_{1/2}$, the $\pi_{1/2}$ series actually starts from a negative $\pi_{1/2}$. The basic fermion formula contains $\pi_{1/2}$'s, S, and N. The only hadronic bond is N – N.

An example is J/ψ ($N_2 S_4$) whose mass (3096.9 ± 0.04 MeV) is calculated as follows.

$$M = 2 M_N + 4 M_S + 4 M_{S-S}$$
$$= 3096.7 \text{ MeV} \tag{32}$$

## 5. *Conclusion*

The periodic table of elementary particles is derived from S-duality between the eleven dimensional membrane and dual lepton-quark superstring (Fig. 1), U-duality between the superstring and solitonic string (Fig. 2), and T-duality between Type II string and compacified string. (Fig. 3). The periodic table of elementary particles (Fig. 4 and Table 2) is constructed from the Kaluza-Klein substructure. All leptons, quarks, and gauge bosons can be placed in the periodic table. The masses of elementary particles can be calculated using only four known constants: the number of the extra spatial dimensions in the eleven dimensional membrane, the mass of electron, the mass of Z°, and $\alpha_e$. The calculated masses (Tables 1 and 3) of elementary particles derived from the periodic table are in good agreement with the observed values. For an example, the calculated mass (176.5 GeV) of the top quark has an excellent agreement with the observed mass (176 ± 13 GeV).

A hadron can be represented by the full quark formula (Table 4) and the basic quark formula (Table 5). The full quark formula consists of all quarks, pseudoscalar quark in pion, and the hybrid quarks. The basic quark formula consists of the lowest mass quarks.   The relation between these two formulas is expressed by Eq. (15).  As a molecule is the composite of atoms with chemical bonds, a hadron is the composite of elementary particles with "hadronic bonds" (Table 6) which are the overlappings of the auxiliary dimensional orbits.   The calculated masses (Tables 7, 8, 9, and 10) of hadrons are in good agreement with the observed values.  For examples, the calculated masses for neutron and pion are 939.54 and 135.01MeV in excellent agreement with the observed masses, 939.57 and 134.98 MeV, respectively.    The overall average difference between the calculated masses and the observed masses for all hadrons (110 hadrons) is 0.29 MeV, and the standard deviation for such differences is 5.06 MeV.  The average observed error for the masses of the hadrons is $\pm$ 6.41 MeV.  The periodic table of elementary particles provides the most comprehensive explanation and calculation for the masses of elementary particles and hadrons.